\documentclass[aps,prx,twocolumn,superscriptaddress,showpacs]{revtex4-1}
\usepackage{amsmath,amssymb,wasysym,graphicx}
\usepackage[utf8]{inputenc}
\usepackage[T1]{fontenc}
\usepackage{xcolor}

\IfFileExists{newtxtext.sty}
{\usepackage{newtxtext,newtxmath}}
{\IfFileExists{stix.sty}
	{\usepackage{stix}}
	{\IfFileExists{mathptmx.sty}
		{\usepackage{mathptmx}}{} } }

\usepackage{textcomp}

\usepackage{bm}

\IfFileExists{siunitx.sty}{\usepackage{booktabs,siunitx}}{}

\usepackage{color}
\definecolor{LinkColor}{rgb}{0.256,0.439,0.588}
\usepackage{hyperref}
\hypersetup{
	pdfauthor={good guys},
	pdftitle={good title},
	colorlinks=true,
	citecolor=LinkColor,
	linkcolor=LinkColor,
	urlcolor=LinkColor
}

\def\normOrd#1{\mathop{:}\nolimits\!#1\!\mathop{:}\nolimits}
\newcommand{\beq} {\begin{equation}}
\newcommand{\eeq} {\end{equation}}
\newcommand{\bea} {\begin{eqnarray}}
\newcommand{\eea} {\end{eqnarray}}
\newcommand{\be} {\begin{equation}}
\newcommand{\ee} {\end{equation}}
\renewcommand{\(}{\left(}
\renewcommand{\)}{\right)}
\renewcommand{\[}{\left[}
\renewcommand{\]}{\right]}

\newcommand{\ket}[1]{\left|#1\right>}
\newcommand{\bra}[1]{\left<#1\right|}

\usepackage{latexsym}
\usepackage{amssymb}
\usepackage{amsmath}
\usepackage{amsfonts}
\usepackage{wasysym}
\usepackage[vcentermath]{youngtab}
\usepackage{upgreek}
\usepackage{bm,dsfont}
\usepackage{multirow}
\usepackage{enumitem}
\usepackage{makecell}
\usepackage{framed}
\usepackage{color}
\usepackage{hyperref}
\usepackage{tikz}
\hypersetup{
colorlinks = true,
linkcolor = [rgb]{0.70,0.13,0.13},
citecolor = [rgb]{0.13,0.55,0.13},
urlcolor = [rgb]{0.25, 0.41, 0.88}}

\def\be{\begin{equation}}
\def\ee{\end{equation}}
\newcommand{\beqn}{\begin{eqnarray}}
\newcommand{\eeqn}{\end{eqnarray}}

\def\Eq#1{Eq.~(\ref{#1})}
\def\Fig#1{Fig.~\ref{#1}}
\def\avg#1{\left\langle#1\right\rangle}

\begin{document}

\title{Exploring nontrivial topology at quantum criticality in a superconducting processor}

\author{Ziqi Tan}\thanks{These authors contributed equally}
\affiliation{College of Computer Science and Technology, Zhejiang University, Hangzhou 310027, China}
\affiliation{School of Physics, ZJU-Hangzhou Global Scientific and Technological Innovation Center, and Zhejiang Key Laboratory of Micro-nano Quantum Chips and Quantum Control, Zhejiang University, Hangzhou 310000, China}

\author{Ke Wang}\thanks{These authors contributed equally}
\affiliation{School of Physics, ZJU-Hangzhou Global Scientific and Technological Innovation Center, and Zhejiang Key Laboratory of Micro-nano Quantum Chips and Quantum Control, Zhejiang University, Hangzhou 310000, China}

\author{Sheng Yang}\thanks{These authors contributed equally}
\affiliation{Institute for Advanced Study in Physics and School of Physics, Zhejiang University, Hangzhou 310058, China}

\author{Fanhao Shen}

\author{Feitong Jin}

\author{Xuhao Zhu}

\author{Yujie Ji}

\author{Shibo Xu}

\author{Jiachen Chen}

\author{Yaozu Wu}

\author{Chuanyu Zhang}

\author{Yu Gao}

\author{Ning Wang}

\author{Yiren Zou}

\author{Aosai Zhang}

\author{Tingting Li}

\author{Zehang Bao}

\author{Zitian Zhu}

\author{Jiarun Zhong}

\author{Zhengyi Cui}

\author{Yihang Han}

\author{Yiyang He}

\author{Han Wang}

\author{Jianan Yang}

\author{Yanzhe Wang}

\author{Jiayuan Shen}

\author{Gongyu Liu}

\author{Zixuan Song}

\author{Jinfeng Deng}

\author{Hang Dong}

\author{Pengfei Zhang}

\affiliation{School of Physics, ZJU-Hangzhou Global Scientific and Technological Innovation Center, and Zhejiang Key Laboratory of Micro-nano Quantum Chips and Quantum Control, Zhejiang University, Hangzhou 310000, China}

\author{Shao-Kai Jian}
\affiliation{Department of Physics and Engineering Physics, Tulane University, New Orleans, Louisiana, 70118, USA}

\author{Hekang Li}

\affiliation{School of Physics, ZJU-Hangzhou Global Scientific and Technological Innovation Center, and Zhejiang Key Laboratory of Micro-nano Quantum Chips and Quantum Control, Zhejiang University, Hangzhou 310000, China}

\author{Zhen Wang}

\affiliation{School of Physics, ZJU-Hangzhou Global Scientific and Technological Innovation Center, and Zhejiang Key Laboratory of Micro-nano Quantum Chips and Quantum Control, Zhejiang University, Hangzhou 310000, China}

\affiliation{Hefei National Laboratory, Hefei 230088, China}

\author{Qiujiang Guo}

\affiliation{School of Physics, ZJU-Hangzhou Global Scientific and Technological Innovation Center, and Zhejiang Key Laboratory of Micro-nano Quantum Chips and Quantum Control, Zhejiang University, Hangzhou 310000, China}

\affiliation{Hefei National Laboratory, Hefei 230088, China}

\author{Hai-Qing Lin}
\affiliation{Institute for Advanced Study in Physics and School of Physics, Zhejiang University, Hangzhou 310058, China}

\author{Chao Song}
\email{chaosong@zju.edu.cn}
\affiliation{School of Physics, ZJU-Hangzhou Global Scientific and Technological Innovation Center, and Zhejiang Key Laboratory of Micro-nano Quantum Chips and Quantum Control, Zhejiang University, Hangzhou 310000, China}

\affiliation{Hefei National Laboratory, Hefei 230088, China}

\author{Xue-Jia Yu}
\email{xuejiayu@fzu.edu.cn}
\affiliation{Department of Physics, Fuzhou University, Fuzhou 350116, Fujian, China}
\affiliation{Fujian Key Laboratory of Quantum Information and Quantum Optics,
College of Physics and Information Engineering,
Fuzhou University, Fuzhou, Fujian 350108, China}

\author{H. Wang}
\email{hhwang@zju.edu.cn}
\affiliation{School of Physics, ZJU-Hangzhou Global Scientific and Technological Innovation Center, and Zhejiang Key Laboratory of Micro-nano Quantum Chips and Quantum Control, Zhejiang University, Hangzhou 310000, China}

\affiliation{Hefei National Laboratory, Hefei 230088, China}

\author{Fei Wu}
\affiliation{College of Computer Science and Technology, Zhejiang University, Hangzhou 310027, China}

\begin{abstract}
The discovery of nontrivial topology in quantum critical states has introduced a new paradigm for classifying
quantum phase transitions and challenges the conventional belief that topological phases are typically associated with a bulk energy gap.
However, realizing and characterizing such topologically nontrivial quantum critical states with large particle numbers remains an outstanding experimental challenge in statistical and condensed matter physics. 
Programmable quantum processors can directly prepare and manipulate exotic quantum many-body states, offering a powerful path for exploring the physics behind these states. 
Here, we present an experimental exploration of the critical cluster Ising model
by preparing its low-lying critical states on a superconducting processor with up to $100$ qubits.
We develop an efficient method to probe the boundary $g$-function based on prepared low-energy states, which allows us to uniquely identify the nontrivial topology of the critical systems under study. 
Furthermore, by adapting the entanglement Hamiltonian tomography technique, we recognize two-fold topological degeneracy in the entanglement spectrum under periodic boundary condition, experimentally verifying the universal bulk-boundary correspondence in topological critical systems.
Our results demonstrate the low-lying critical states as useful quantum resources for investigating the interplay between topology and quantum criticality.
\end{abstract}

\maketitle

\section{Introduction}
Throngs of simple interacting elements can self-organize to exhibit rich emergent phenomena. In quantum many-body systems, this principle of emergence explains various exotic properties of quantum matter ranging from superconductivity and magnetism to topological states, which are essential for modern quantum devices and technologies.
For decades, tremendous efforts have been devoted to recognizing these intrinsic orders of quantum matter, with the understanding advanced from the traditional Landau-Ginzburg-Wilson symmetry-breaking paradigm to topological phases which are rooted in many-body entanglement~\cite{sachdev1999quantum,hasan2010rmp,qi2011rmp,wen2017rmp}. Typically, topological phases are associated with a bulk energy gap, and it has long been believed that nontrivial topological properties are destroyed when the gap closes. 
However, recent progress has opened the possibility of nontrivial topology in gapless systems, leading to the notion of gapless symmetry-protected topological (gSPT) phases~\cite{scaffidi2017prx,verresen2018prl,verresen2021prx, yu2022prl}.
A pioneering example on this front involves the topologically nontrivial quantum critical points (QCPs), which has garnered considerable attention in recent years~\cite{thorngren2021prb,duque2021prb,yu2024prl,yu2024prb,yang2024giftslongrangeinteractionemergent,zhong2024pra,li2023intrinsicallypurelygaplesssptnoninvertibleduality,huang2023topologicalholographyquantumcriticality,wen2023prb,wen2023classification11dgaplesssymmetry,li2024sci_post,zhang2024pra}.
In addition, the discovery of nontrivial topology at critical points not only broadens the scope of topological physics to more challenging gapless many-body systems, but also opens new avenues for classifying phase transitions within the same universality class, fundamentally enriching the textbook understanding of phase transitions.

\begin{figure*}[!t] \begin{center}
    \includegraphics[width=\textwidth]{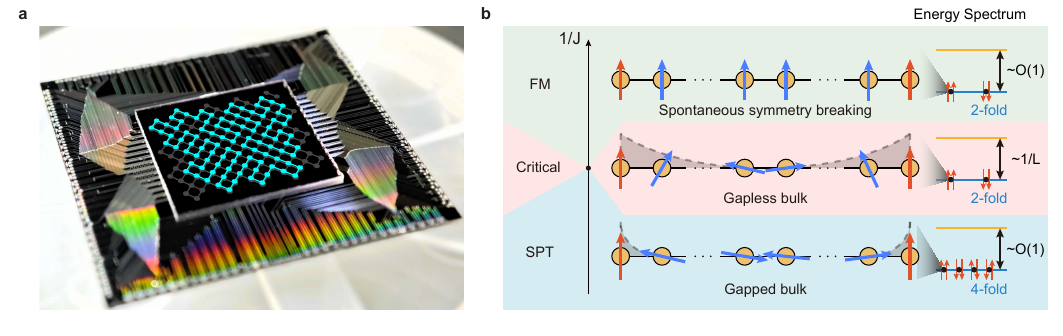}

    \caption{
    \textbf{Devices and models.} \textbf{a}, A photograph of the superconducting quantum processor. The qubit layer and control-line layer are patterned on the top and bottom substrates respectively, which are assembled together during the flip-chip bonding process. The qubits used in this work, which can be connected to form a one-dimensional ring, are highlighted in cyan. \textbf{b}, Schematic representation of the cluster Ising chain with open boundary conditions, which undergoes a phase transition between the ferromagnetic (FM) long-range order and the cluster SPT phase. The gapless SPT state is located at the critical point, featuring localized zero-energy edge states and two-fold degeneracy. The critical bulk gap scales inversely proportional to the system size $L$, which vanishes at the thermodynamic limit.
    }
    \label{fig:fig1}
\end{center} \end{figure*}

However, the investigation of quantum criticality is challenging due to the complex entanglement induced by quantum fluctuations on all length scales.
As a result, the experimental detection of topology at QCPs remains elusive.
Recent advances of programmable quantum processors across various physical platforms~\cite{Google2024Quantum, Kim2023_Evidence, Xu2023nonAbelian, Bluvstein2024_logical, Iqbal2024_nonAbelian} have opened a new path to explore exotic phases of quantum matter by directly engineering the quantum many-body states of interest.
To efficiently harness the quantum resources, quantum variational techniques have been developed~\cite{McClean2016_theTheory, Kokail2019_selfVerifying, Arute2020_Hartree}.
In this approach, a family of variational trial states $|\Psi_v(\boldsymbol{\theta})\rangle$, parameterized by a vector $\boldsymbol{\theta}$, is optimized to approximate the low-energy state of a target Hamiltonian $H$.
It relieves the quantum processors from directly realizing the Hamiltonian and endows great flexibility for quantum simulation.
However, when restricted to the goal of preparing the ground state, the scalability of the method is often questioned due to the existence of barren plateaus and local minima in the optimization landscape~\cite{larocca2024_review, McClean2018_Barren, Bittel2021_training}.
For gapless quantum many-body systems, additional challenges arise as the energy of the system is no longer a good objective function.
In contrast, the potential of variationally prepared low-lying states, which are experimentally more accessible, remains relatively underexplored.

Here, we demonstrate how the low-lying quantum states can be harnessed to detect key properties of the gSPT phases.
We perform quantum simulations to probe the nontrivial topological properties of the critical cluster Ising model, using a superconducting quantum processor with up to 100 qubits. By devising variational quantum circuits that respect the symmetries of the target Hamiltonian, we present an efficient way to generate the low-lying critical states at the QCPs with a predominant ground state component.
Based on the prepared states, we address the challenges of extracting the topological properties of the QCPs under study.
We first focus on the Affleck-Ludwig boundary $g$-function, which serves as the topological invariant of the quantum criticality~\cite{verresen2018prl,verresen2021prx,yu2022prl,yu2024prl}.
We propose and experimentally verify an efficient method to measure the boundary $g$-function based on the overlaps between low-energy wavefunctions of different boundary conditions.
We then turn to the detection of the entanglement spectrum, which contains universal information that goes beyond the entanglement entropy in characterizing gapless topological phases.
We adapt the entanglement Hamiltonian tomography (EHT) technique~\cite{Kokail2021_entanglement} to probe the entanglement spectrum of the quantum critical states and obtain results that agree decently well with the theory.
Our results provide the first experimental observation of nontrivial topology at QCPs, suggesting widely accessible low-lying many-body states as useful quantum resources for quantum simulation.

\section{Experimental setup and model}
 {The experiment is carried out on a flip-chip superconducting quantum processor featuring a rectangular grid of 125 frequency-tunable transmon qubits, as shown in Fig.~\ref{fig:fig1}a.
Each qubit has an individual control line for realizing arbitrary single-qubit gates, and is coupled to an individual resonator for dispersive quantum state readout.
The nearest neighboring qubits are connected by a tunable coupler, which enables the implementation of two-qubit CZ gates~\cite{Ren2022_experimental}.
For the 100 qubits selected in this experiment, we realize median fidelities above $99.94\% (99.57\%)$ and $99.14\%$ for the parallel single- (two-) qubit gates and readout, respectively.
The configuration of our device supports a flexible construction of the one-dimensional (1D) qubit chain with different boundary conditions.
}

 {To study the universal properties of the gapless topological phases, we focus on realizing the low-lying critical states of the cluster Ising (CI) model in a 1D spin chain, described by the Hamiltonian:
\begin{equation}
    \label{E1}
    H_{\mathbb{Z}_{2}} = - \sum_{j} Z_{j} Z_{j+1} -J \sum_{j}Z_{j-1} X_{j} Z_{j+1}\,,
\end{equation}
where $X_{j}$ and $Z_{j}$ represent the Pauli matrices on each site $j$, and $J$ parameterizes the cluster interaction strength.
The model exhibits a $\mathbb{Z}_{2}$ spin-flip symmetry (generated by $O_X = \prod_{i} X_{i}$) and a time-reversal symmetry $\mathbb{Z}_{2}^{T}$ (acting as complex conjugation). 
The cluster interaction drives the system toward the $\mathbb{Z}_{2} \times \mathbb{Z}_{2}^{T}$ SPT phase, which is sometimes referred to as the cluster or Haldane SPT phase~\cite{Son_2011,verresen2017prb}.

As $J$ varies, the model undergoes a phase transition between the ferromagnetic long-range order and the cluster SPT phase.
The location of the critical point is pinned at $J=1.0$ due to self-duality property~\cite{kogut1979rmp}.
The bulk critical behaviors of the critical cluster Ising model are the same as the familiar critical transverse field Ising model, both characterized by the Ising conformal field theory (CFT)~\cite{ginsparg1988appliedconformalfieldtheory}. 
However, the boundary physics of the model can exhibit additional features~\cite{binder1974prb,CARDY1984514}.
Specifically, the QCP of the cluster Ising model is topological nontrivial, as it supports localized zero-energy edge states and spontaneously breaks $\mathbb{Z}_{2}$ spin-flip symmetry near the boundaries~\cite{verresen2018prl,verresen2021prx,yu2022prl,yu2024prl}. Therefore, the QCP for the critical cluster Ising model is topologically distinct from the critical transverse field Ising model even though they belong to the same universality class, which provides the simplest example of the gSPT state (see Supplementary Materials I.A for a general introduction).
Preparing the low-energy states at this point with both periodic and open boundary conditions allows us to experimentally detect the topological invariant and verify the bulk-boundary correspondence of the gapless topological phases.

\begin{figure*}[!t] \begin{center}
    \includegraphics[width=\textwidth]{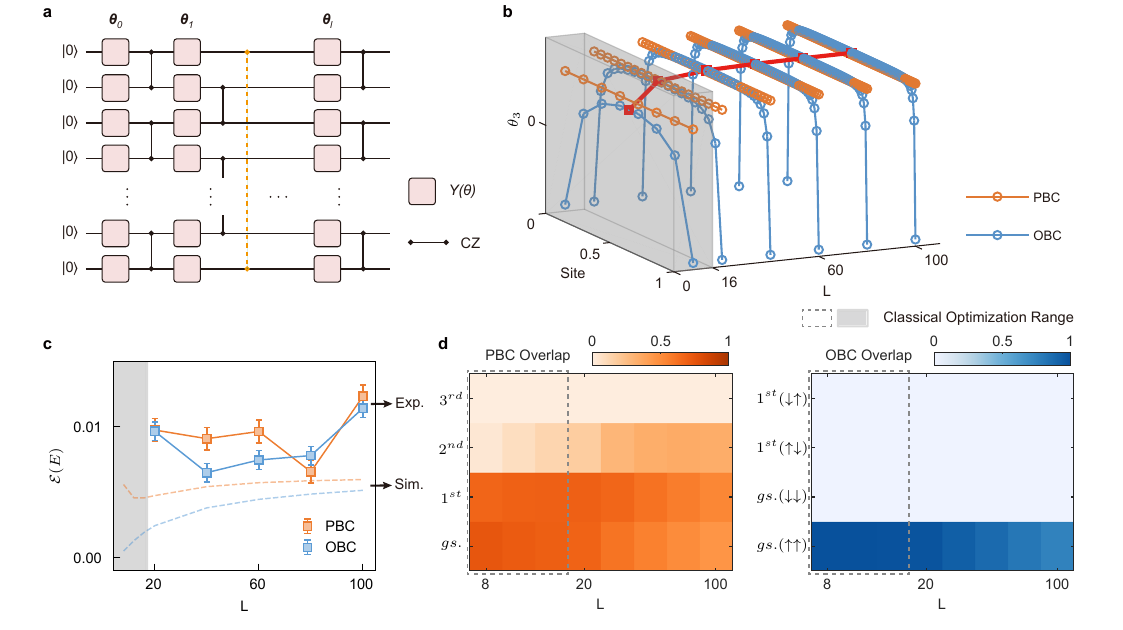}
        
    \caption{\textbf{Scalable preparation of the low-lying critical states with variational quantum circuits.} \textbf{a}, The variational ansatz used in this work, which consists of five alternating variational and entangling layers. Each variational layer is composed of single-qubit $Y(\theta)$ gates. The entangling layer can be constructed with two different patterns of two-qubit CZ gates, which are applied alternatively.
    The CZ gates highlighted by yellow are additionally applied for preparing states under PBC.
    \textbf{b}, The parameters of the third variational layer for different system sizes. For each fixed system size, the parameters share a single value for PBC, and follow a power-law decay with the increase of the site distance to the boundary for OBC.
    The shaded area refers to the regime where the variational parameters are obtained using numerical simulation and optimization on a classical computer. 
    For systems with larger sizes, we apply a power-law extrapolation to directly obtain the parameters.
    \textbf{c}, Normalized energy distances between the prepared low-lying critical states and the ground states, defined as $\varepsilon(E)=\frac{E-E_g}{E_\text{max}-E_g}$, with different boundary conditions and system sizes.
    Here, $E$ is the energy of the low-lying state measured experimentally with error mitigation, $E_g$ denotes the theoretical ground state energy, and $E_\text{max}$ denotes the theoretical maximum eigenenergy. 
    Dashed lines denote numerical simulations, and square dots are the experimentally measured data. Error bars represent the standard deviation obtained from bootstrapping.
    \textbf{d}, The numerical analyses of the overlaps between the prepared low-lying critical states and the lowest four eigenstates of the critical CI model under OBC and PBC.
    } \label{fig2}
\end{center} \end{figure*}

\section{Preparing the low-lying critical states}

We experimentally prepare the low-lying critical states with variational quantum circuits determined solely by a classical computer, avoiding the challenges of the traditional hybrid quantum-classical variational methods.
Since the target states have real-valued amplitudes, we construct the ansatz circuits with interleaved layers of variational single-qubit $Y(\theta)$ gates (rotation around the $Y$ axis by $\theta$) and two-qubit CZ gates (Fig.~\ref{fig2}\textbf{a}).
The resulting trial states can be expressed as $|\Psi_v(\boldsymbol{\theta})\rangle=U(\boldsymbol{\theta})|0\rangle^{\otimes L}$, with $L$ being the qubit number.
We model the variational parameters in each single-qubit gate layer with a site-dependent parameter function, as exemplified in Fig.~\ref{fig2}\textbf{b},  which incorporates the symmetries of the model with either periodic or open boundary conditions.
We optimize the parameter function by maximizing the fidelities with the target states, using systems of up to 16 qubits.
We then directly extrapolate the circuit to large system sizes with up to 100 qubits.
See Supplementary Materials III.A for details.

We apply the method to prepare the low-lying critical states of the critical CI
model with different boundary conditions.
The resulting states have energy distances from the corresponding ground states of less than $1.5\%$ of the entire spectral ranges, as verified by both numerical simulations and experimental measurements (Fig.~\ref{fig2}\textbf{c}). 
To gain a deeper understanding of the prepared states, we numerically analyze their components based on matrix product state (MPS)~\cite{cirac2006prb} circuit simulation and density matrix renormalization group (DMRG) calculation~\cite{white1992prl}.
In Fig.~\ref{fig2}\textbf{d}, we display the overlaps between the variationally prepared states and each of the lowest four energy eigenstates of the critical CI model under open and periodic boundary conditions for different system sizes.
For the open boundary condition (OBC), the low-lying critical states feature a dominant ground state component, with an overlap beyond 0.74 for system sizes up to 100 qubits.
For the periodic boundary condition (PBC), the low-lying critical states have two dominant components, namely the lowest two energy eigenstates, each of which has an overlap beyond 0.45\,.

\begin{figure}[!t] \begin{center}
    \includegraphics[width=\columnwidth]{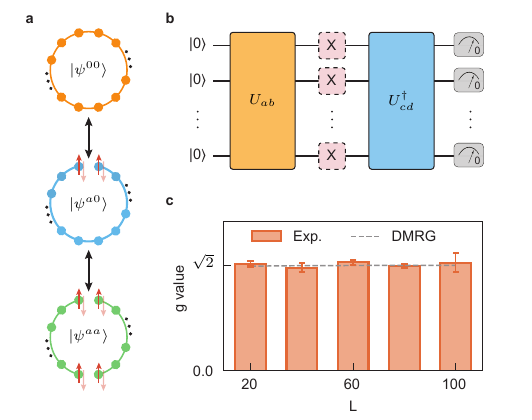}
        
    \caption{\textbf{Detecting the boundary $g$-function.} \textbf{a}, The three configurations of 1D spin chain used in measuring the boundary $g$-function with the state overlap method, which are realized by applying zero, one, and two physical cuts to the original spin chain under PBC. The corresponding ground states are denoted as $|\psi^{ab}\rangle$, with $a$ and $b$ marking the physical boundary conditions on the site pairs $(0, L-1)$ and $(L/2-1, L/2)$, respectively.
    In this work, $a$ and $b$ are taken from $\{0, \uparrow, \downarrow\}$, which represent the physical boundary conditions of no cut ($0$) and one cut with both spins fixed at spin up ($\uparrow$) or spin down ($\downarrow$) states. \textbf{b}, Measurement scheme for the overlap between the low-lying critical states of two different boundary conditions. 
    $U_{ab (cd)}$ represent the variational circuit for preparing the state with boundary condition denoted by $ab (cd)$, i.e., $U_{ab (cd)}|0\rangle^{\otimes L}=|\psi^{ab (cd)}\rangle$.
    The $X$ gates are applied in-between alternatively for state projection.
    \textbf{c}, Experimentally measured $g$-functions of the critical CI model with different system sizes $L$, with error bars denoting the standard deviation from five repetitions of measurements. The dotted line denotes the ideal $g$-function obtained by DMRG. } \label{fig3}
\end{center} \end{figure}

\section{Detecting the boundary $g$-function}
With the low-lying critical states prepared, we now show how they can be leveraged to detect the topological properties of the critical systems.
The first quantity we consider is the boundary $g$-function, which can serve as a topological invariant at criticality~\cite{yu2022prl} (a more detailed discussion can be found in Supplementary Materials I.B).
Theoretically, 
the boundary $g$-function can be calculated with the overlaps between ground states of spin chains under different boundary conditions: $g=\sum_{a\in\{\uparrow, \downarrow\}}\left|\frac{\langle\psi^{a0}|\psi^{00}\rangle}{\langle\psi^{a0}|\psi^{aa}\rangle}\right|$~\cite{Zhou2024SciPost}. 
Here $|\psi^{ab}\rangle$ is the ground state with the boundary condition $\{a,b\}$ realized by the physical cuts on the site pairs $(0, L-1)$ and $(L/2-1, L/2)$ (see Fig.~\ref{fig3}\textbf{a}).
The notation ‘0’ means doing nothing on the site pair.

Our method extends the state overlap approach by representing the ground states approximately with the low-lying critical states of the same kind of boundary conditions.
The substitution of the ground states with OBC is natural since they are the dominant components in the corresponding low-lying critical states.
For the low-lying critical states with PBC, we project out the constituent eigenstates that are outside the $\mathbb{Z}_{2}$ symmetry sector during the overlap calculation, which include the first excited state.
As a result, we obtain the $g$-function as $g=\sum_{a\in\{\uparrow, \downarrow\}}\left|\frac{\langle\Psi_v^{a0}|\Psi_{v}^{00,+}\rangle}{\langle\Psi_v^{a0}|\Psi_v^{aa}\rangle}\right|$, where $|\Psi_{v}^{00,+}\rangle=\frac{\mathcal{P}}{\sqrt{|\langle\Psi_v^{00}|\mathcal{P}|\Psi_v^{00}\rangle|}}|\Psi_v^{00}\rangle$ with $\mathcal{P}=(I+O_X)/2$ being the projector which projects $|\Psi_v^{00}\rangle$ into the $\mathbb{Z}_{2}$ symmetric sector. 

The quantum circuit for measuring the overlap between two real-valued low-lying critical states is shown in Fig.~\ref{fig3}\textbf{b}, which consists of $U_{ab}$ and $U_{cd}^\dagger$, i.e., the inverse of $U_{cd}$, followed by measuring the probability of the system in the initial state $|0\rangle^{\otimes L}$. Here, $U_{ab (cd)}$ denotes the variational circuit for
preparing $|\Psi_v^{ab (cd)}\rangle$.
According to Born's rule, the measured probability gives the overlap between two low-lying critical states: $|\langle\Psi_v^{ab}|\Psi_v^{cd}\rangle|=\sqrt{P}$.
The $X$ gate layer in-between is applied alternatively for realizing the symmetry projection. See Supplementary Materials III.B for details.

In Fig.~\ref{fig3}\textbf{c}, we plot the experimentally measured values of $g$-function, which remain at the nontrivial value of around $\sqrt{2}$ for different system sizes due to the fixed conformal boundary condition and agree well with the theoretical results obtained with DMRG calculation. 
We attribute the robustness of the results to the fact that the $g$-function can also be extracted from the low-energy eigenstates, given that they are in the $\mathbb{Z}_{2}$-even sector (see the physical derivation in the Supplementary Materials I.C).
In addition, the experimental errors partially cancel out during the division of the overlaps, making the method intrinsically robust against the imperfections of quantum gates and measurements. In comparison, the conformal boundary condition of the critical transverse field Ising model is known as the free boundary condition, resulting in a boundary $g$-function equal to $1$ and belonging to the topological trivial Ising universality class. Therefore, the apparent departure of the experimentally measured $g$-function from $1$ provides solid and compelling evidence of the nontrivial topological properties at the QCPs.

\section{Detecting the entanglement spectrum}
Now we turn to the entanglement spectrum, which stands as another key quantity in characterizing the gapless topological phase~\cite{yu2024prl}.
The entanglement spectrum encodes information about both the topological degeneracy and the corresponding boundary CFT operator content, extending beyond what is captured by entanglement entropy.
For the CI model considered in this work, the two-fold topological degeneracy at the critical point can be directly revealed without imposing open boundaries, through the well-known Li-Haldane bulk-boundary correspondence in the entanglement spectrum under PBC~\cite{Li2008PRL,pollmann2010prb} (see Supplementary Materials I.D for details). 
Using entanglement Hamiltonian tomography (EHT) techniques~\cite{Kokail2021_entanglement}, we show that the low-lying critical states provide useful resources for recovering the reduced density operators of subsystems in the ground state with high fidelities, from which the entanglement spectrum can be revealed. 

At the heart of EHT is the construction and learning of the entanglement Hamiltonian (EH) $\widetilde{H}_A$, which is connected to the reduced density matrix of a subsystem $A$ via $\rho_A\approx e^{-\widetilde{H}_A}$.
By adapting the Bisognano–Wichmann theorem~\cite{bw1975a} from the quantum field theory to discrete lattice models, we construct the EH ansatz as following:
\begin{equation}
\label{E2}
    \begin{gathered}
        \widetilde{H}_{A} = -\sum_{j,j+1\in A} \beta_j^{ZZ}Z_{j} Z_{j+1} - \\
        \sum_{j,j\pm1\in A} \beta_j^{ZXZ}Z_{j-1} X_{j} Z_{j+1} -\sum_{j\in A}\beta_j^{X}X_{j} \,,
    \end{gathered}
\end{equation}
where $\beta_j^{ZZ}$, $\beta_j^{ZXZ}$, and $\beta_j^{X}$ are parameters which characterize the inverse entanglement temperature profile.
The EH ansatz can be seen as the local deformation of the system Hamiltonian Eq.(\ref{E1}) that exhibits the same $\mathbb{Z}_{2}$ symmetry.
With efficient parameterization of the EH, the EHT technique realizes an exponential reduction of the measurement bases required for reconstructing the reduced density matrix compared with conventional quantum state tomography.
More importantly, the approach remains robust in unveiling the entanglement structure of the ground state from the low-lying critical states of the system.

\begin{figure}[!t] 
\begin{center}
    \includegraphics[width=\columnwidth]{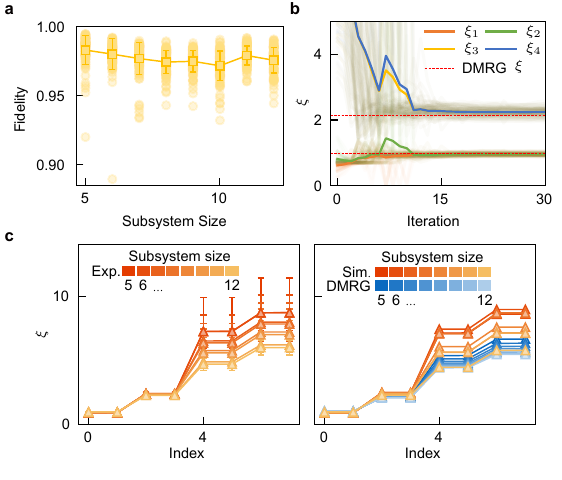}
        
    \caption{\textbf{Detecting the entanglement spectrum.} \textbf{a}, Fidelities of the reconstructed density matrices for different subsystem sizes. For each subsystem size, we reconstruct the density matrices of 100 different choices of subsystems in the 1D chain, and calculate the fidelity as $F=\left(\text{Tr}\sqrt{\sqrt{\rho_\text{exp}}\cdot\rho_\text{DMRG}\cdot\sqrt{\rho_\text{exp}}}\right)^2$, where $\rho_\text{exp}$ and $\rho_\text{DMRG}$ denotes the density matrix obtained with EHT and DMRG, respectively. Transparent circles show the fidelities of all density matrices. Square dots and error bars denote the mean and standard deviation of the 100 calculated fidelities.
    \textbf{b}, Entanglement spectrum of the subsystem with $8$ qubits during the EHT learning procedure. \textbf{c}, Entanglement spectrum of subsystems with different sizes obtained from the experimentally prepared low-lying states (left). We numerically simulate the state preparation and EHT procedure without considering experimental noises, with the results shown in the right panel. As a comparison, we also display the entanglement spectrum obtained directly from the numerically calculated ground states. The lowest eight levels $\xi$ are displayed. The experimental data is averaged over 100 different subsystems from the 1D chain, with the error bars denoting the standard deviation.} \label{fig4}
\end{center} 
\end{figure}

We experimentally implement EHT to analyze the entanglement spectrum of the bulk of the critical CI model with up to 100 qubits.
After preparing the low-lying state of the critical CI model under PBC, we measure the system under {200} Pauli bases and in each base collect 3,000 samples from quantum projective measurements, from which we learn the EHs for subsystems with sizes ranging from $5$ to $12$.
For each size, we iterate on the 1D chain to extract the reduced density matrices of $100$ different choices of subsystems.
We verify the accuracy of the learned EH by comparing the reconstructed density matrices with the theoretical ones obtained by DMRG.
As shown in Fig.~\ref{fig4}\textbf{a}, the fidelities of the learned density matrices for all subsystems throughout the chain are centered around $0.97$.
In Fig.~\ref{fig4}\textbf{b}, we plot the entanglement spectrum of subsystems with $8$ qubits at different stages of the EHT data learning procedure.
As expected, the two-fold degeneracy is gradually recovered from the experimental data.
The recovered entanglement spectrum for all subsystem sizes is summarized in Fig.~\ref{fig4}\textbf{c}, where the two-fold degeneracy is observed in all cases, agreeing well with numerical simulation.
The results provide the first experimental evidence of the Li-Haldane conjecture for the gapless SPT state~\cite{yu2024prl}, confirming the validity of low-lying critical states in exploring many-body physics. In Supplementary Materials I.D, we provide a simple argument to explain the robust double topological degeneracy observed in the bulk entanglement spectrum of the low-lying critical state.

\section{Conclusion and outlook}
In summary, we have performed quantum simulation of a critical cluster Ising chain and observed to our knowledge the first compelling signatures of the nontrivial topology at quantum critical points with up to 100 qubits, pushing forward the experimental exploration of gapless SPT phases and demonstrating the utility of the widely accessible low-lying states.
Interestingly, both quantities measured in this work, i.e. the boundary $g$-function and the entanglement spectrum, are related to the topological properties of the critical system, indicating the robustness of these properties against both perturbations in the quantum critical state and experimental imperfections.
While ground states are often used in literature to study quantum matter, it would be interesting and important to extend the investigation to low-energy states.
Indeed, encouraging findings that low-energy states may also encode universal information of quantum matter have recently been reported~\cite{Zou2022PRB,Zhou2024SciPost}.
As we venture towards achieving quantum simulation beyond classical capability, our findings could open new avenues for future explorations of fundamental physics such as quantum criticality in higher dimensions, avoiding the task of preparing large-scale ground states which could be prohibitively challenging for noisy quantum devices.

\textit{Acknowledgement}: We thank Yi-Zhuang You for helpful discussions. The device was fabricated at the Micro-Nano Fabrication Center of Zhejiang University. We acknowledge the support from the Innovation Program for Quantum Science and Technology (Grant No. 2021ZD0300200), the National Natural Science Foundation of China (Grant Nos. 92365301, 12174342, 12405034, 12274368, 12274367,
12322414, 12404570, 12404574, U20A2076), the National Key Research and Development Program of China (Grant No. 2023YFB4502600), and the Zhejiang Provincial Natural Science Foundation of China (Grant Nos. LR24A040002 and LDO23A040001).

\bibliography{main}

\clearpage
\onecolumngrid

\newpage
\begin{widetext}

\def\normOrd#1{\mathop{:}\nolimits\!#1\!\mathop{:}\nolimits}
\renewcommand{\(}{\left(}
\renewcommand{\)}{\right)}
\renewcommand{\[}{\left[}
\renewcommand{\]}{\right]}

\setcounter{figure}{0}
\setcounter{section}{0}
\renewcommand{\thefigure}{S\arabic{figure}}

\def\Eq#1{Eq.~(\ref{#1})}
\def\Fig#1{Fig.~\ref{#1}}
\def\avg#1{\left\langle#1\right\rangle}

\begin{center} 
	{\large \bf Supplementary Materials: Exploring nontrivial topology at quantum criticality in a superconducting processor}
\end{center}

\section{Theoretical Understanding}

\subsection{Introduction to the gapless symmetry-protected topological phases}

To provide a more intuitive understanding of symmetry-protected topological properties in critical systems, we first introduce the fundamental concepts of gapless symmetry-protected topological phases (gSPTs).
In this work, we study a prototypical model called the critical cluster Ising model in experiments as a concrete example.

Topological phases form a cornerstone of modern condensed matter physics, extending beyond the Landau-Ginzburg-Wilson paradigm of symmetry-breaking. 
A notable example of topological states is symmetry-protected topological phases~\cite{wen2017rmp,qi2011rmp,hasan2010rmp}, which are typically associated with a bulk energy gap. 
It has been widely believed that these topological properties are destroyed when the bulk gap closes. 
However, recent achievements~\cite{scaffidi2017prx,verresen2018prl,verresen2021prx,yu2022prl,yu2024prl} have shown that many key features of topological physics, such as topological edge modes, can surprisingly emerge in the less explored realm of gapless quantum many-body systems.
These systems include stable gapless phases or critical points, leading to the notion of gSPT phases. 

To date, different families of gSPTs have been identified in the literature~\cite{li2023intrinsicallypurelygaplesssptnoninvertibleduality,wen2023classification11dgaplesssymmetry,wen2023prb,li2024sci_post}. 
And a classification of gSPT phases based on whether they contain gapped sector and whether they are intrinsic is proposed, thereby leading to the categorization into four distinct types, as summarized in Table~\ref{tab:gSPT}. 
The subsequent detailed explanations are mainly based on the literature~\cite{li2023intrinsicallypurelygaplesssptnoninvertibleduality}. 

\begin{table}[h]
\centering
\begin{tabular}{|c|c|c|}
    \hline
    & Contains gapped sector & No gapped sector  \\
    \hline
    Non-intrinsic  & gSPT  & purely gSPT \\
    \hline
    Intrinsic & igSPT  & intrinsically purely gSPT\\
    \hline
\end{tabular}
\caption{Classification of the gSPTs by whether they are purely gapless (horizontal direction) or intrinsically gapless (vertical direction). 
The table is taken from Ref.~\cite{li2023intrinsicallypurelygaplesssptnoninvertibleduality}.}
\label{tab:gSPT}
\end{table}

i) \emph{ The gSPT phases}: The first example of a gSPT phase, which is non-intrinsic and includes a gapped sector, was systematically investigated in Ref.~\cite{scaffidi2017prx}. 
Specifically, a general construction of gSPT phases was introduced based on the decorated domain-wall (defect) picture of gapped SPT phases~\cite{chen2014symmetry}. 
The central idea is to decorate the $G$-symmetry defects of a $G$-symmetric gapless system or conformal field theory (CFT) with an $H$-symmetric gapped SPT. 
This decorated defect construction creates a gapped sector that acts both on $G$ and $H$ symmetry, resulting in a gSPT phase whose topological properties can also manifest in gapped counterparts. 
As a result, these features are not ``intrinsic'' to the gapless system~\cite{scaffidi2017prx,li2024sci_post}. 
In summary, the topological properties of these gSPT phases arise from the gapped SPT sector and can be interpreted as a gapped SPT stack with a CFT. 
Recent studies~\cite{verresen2021prx} have shown that such gSPT states can emerge at conformally invariant critical points separating spontaneously symmetry-breaking (SSB) phases and gapped SPT phases, also known as symmetry-enriched CFT or topologically nontrivial quantum critical points~\cite{verresen2021prx}. 

ii) \emph{The intrinsically gSPT phases}: On a different front, there exists an intriguing class of gapless topological phases, referred to as intrinsically gSPT (igSPT) phases~\cite{thorngren2021prb}, whose topological features are fundamentally prohibited in gapped counterparts. 
Specifically, recent works~\cite{li2023intrinsicallypurelygaplesssptnoninvertibleduality,li2024sci_post} propose a systematic construction of igSPT phases, which we briefly outlined below: The total symmetry group is denoted by $ \Gamma $, fitting into the group extension $ 1 \to H \to \Gamma \to G \to 1 $. 
The construction begins with a $ G $-symmetric gapless system or CFT characterized by a self-anomaly $ \omega_G $. 
An $H$-symmetric gapped SPT phase is then stacked on top of the $G$ symmetry domain walls (defects). 
Due to the non-trivial group extension, the resulting gapped sector exhibits an (emergent) anomaly $ -\omega_G $ that cancels the anomaly in the gapless sector, rendering the combined system an igSPT phase that is $ \Gamma $-anomaly-free. 
By construction, igSPT phases also include a gapped sector, leading to exponentially localized edge modes near the boundaries. 
Importantly, the topological features of igSPT phases cannot be realized in any $ \Gamma $-symmetric gapped SPT phase, thereby justifying the term ``intrinsic''. 
Furthermore, Li et al.~\cite{li2023intrinsicallypurelygaplesssptnoninvertibleduality,li2024sci_post} utilized the decorated defect construction and the Kennedy-Tasaki (KT) transformation to construct analytically tractable 1+1D spin models of both gSPT and igSPT phases, focusing on $ \mathbb{Z}_2 \times \mathbb{Z}_2 $ symmetry and $ \mathbb{Z}_4 $ symmetry, respectively. 
Additionally, the recently developed topological holography principle (known as symmetry topological field theory) was employed to provide a unified classification of gapped and gapless SPT phases from a new perspective~\cite{huang2023topologicalholographyquantumcriticality,wen2023classification11dgaplesssymmetry}. Experimentally, these igSPT phases can emerge at the transition point between a quantum spin Hall insulator and an $s$-wave superconducting phase, which has recently been proposed to be realizable in materials such as WTe$_2$~\cite{song2024unconventional}.

iii) \emph{The purely and intrinsically purely gSPT phases}: From the previous discussion, we have established that both gSPT and igSPT phases typically include a gapped sector, which results in exponentially localized topological edge modes. However, Verresen et al.\cite{verresen2017prb,verresen2021prx} studied a cluster Ising model with time-reversal symmetry and demonstrated that this model lacks a gapped sector, as evidenced by the algebraically decaying energy splitting of edge modes that are forbidden in gapped systems. 
This result suggests the existence of a gSPT phase without any gapped sector, referred to as a ``purely gSPT phase''~\cite{wen2023classification11dgaplesssymmetry,wen2023prb,li2023intrinsicallypurelygaplesssptnoninvertibleduality,li2024sci_post}.
However, to the best of our knowledge, the ground state of the critical cluster Ising chain provides the only known example of a purely gSPT phase.
Therefore, constructing additional lattice Hamiltonians that realize such novel phases is highly desirable. 
Furthermore, it is natural to explore the possibility of an igSPT phase without any gapped sectors, which has been termed an ``intrinsically purely gapless SPT phase''~\cite{li2023intrinsicallypurelygaplesssptnoninvertibleduality}. 
Unfortunately, despite ongoing efforts~\cite{li2023intrinsicallypurelygaplesssptnoninvertibleduality}, constructing such an intrinsically purely gSPT phase remains an open challenge, leaving it as an important direction for future research.

It is important to emphasize that, although a topological semimetal can be broadly considered a gapless SPT phase, there is a qualitative distinction between them. The key difference lies in their topological properties: the former relies on space-translational symmetry, making it susceptible to destabilization by disorder. In contrast, the topological edge modes at quantum criticality remain robust even in the presence of symmetry-preserving disorder.

To provide a concrete and experimentally accessible example, we introduce the following prototypical lattice Hamiltonian, previously studied in Ref.~\cite{duque2021prb,yu2024prb}, as an example of a non-intrinsic purely gSPT phase: 
\begin{equation}\label{eq: Jhg model}
H = -\sum_{i}J Z_{i-1}X_{i} Z_{i+1}-\sum_{i}g Z_i Z_{i+1}-\sum_{i}h X_i \,,
\end{equation}
where $X,Y,Z$ denote the Pauli matrices. The model enjoys the $\mathbb{Z}_2$ spin-flip symmetry (generated by $P = \prod_i X_i$) and the time-reversal symmetry $\mathbb{Z}_2^T$ (acting as the complex conjugation). 
The parameters $g,h, J$ represent ferromagnetic (FM) coupling, transverse field, and cluster interactions, respectively, which drive the system toward FM, trivial paramagnetic (PM), and $\mathbb{Z}_2\times\mathbb{Z}_2^T$ SPT~\cite{Son_2011,verresen2017prb} phases, respectively. When $h = 0, g =1$, the model reduces to the cluster Ising model discussed in the main text.

We notice that although the FM-PM ($J=0, g=h$) and FM-SPT ($h=0, J=g$) transitions are both described by the 1+1D Ising CFT, the time-reversal symmetry acts differently on the disorder operator, leading to different symmetry-enriched CFTs~\cite{verresen2021prx}. 
We refer to the former as the topologically trivial and the latter as the topologically nontrivial Ising critical point. 
To briefly explain why this topological distinction arises, we note that 1+1D Ising CFT has a unique local (nonlocal) scaling operator with scaling dimension $\Delta = 1/8$, typically denoted by $\sigma$ ($\mu$). 
These serve as the order parameters of the adjacent phases: $\sigma(i) \sim Z_i$ is the Ising order parameter, while the disorder operator $\mu(i)$ is the Kramers-Wannier-dual string order parameter that characterizes the symmetry-preserving phase. 
In the trivial PM phase, $\mu(i) \sim \prod_{j=-\infty}^i X_j$, while in the cluster SPT phase, $\mu(i) \sim ( \prod_{j=-\infty}^{i-1} X_{j} ) Y_{i} Z_{i+1}$~\cite{verresen2017prb}. 
The distinction is reflected in the symmetry charge of the disorder operator under time-reversal symmetry: $T \mu T = \pm \mu$, which means that the two critical points can not be smoothly connected and must be separated by another phase transition. 
In the non-trivial case ($T \mu T = - \mu$), a topologically protected edge mode persists even when the bulk gap closes~\cite{verresen2021prx}. 
Intuitively, the boundary of such a critical system spontaneously breaks the $\mathbb Z_2$ spin-flip symmetry, resulting in a two-fold degenerate edge mode that is stable because the charged $\mu$ operator cannot condense near the boundary. 
The finite-size splitting of this edge mode decays algebraically as $\sim 1/L^{14}$, which is parametrically faster than the finite-size bulk gap $\sim 1/L$. 
This faster decay ensures the stability of the topological edge mode, even in the presence of critical bulk fluctuations. 
In essence, the topologically distinct Ising critical points realize different conformal boundary conditions and boundary $g$-functions at low energy, serving as a form of ``generalized ground-state degeneracy'' even for gapless topological phases that lack edge modes~\cite{yu2022prl}.

\subsection{The boundary $g$-function serves as a topological invariant for classifying topologically distinct quantum critical states} \label{sec:theory_g}

Traditionally, people are often interested in the low-energy universal behavior of critical many-body systems in the thermodynamic limit, known as the universality class or more specifically, the bulk universality class.
This is typically described by the (bulk) CFT. 
A key quantity in determining such bulk critical universality is the central charge, which can be extracted from the entanglement entropy of a subsystem of length $l$ under periodic boundary conditions (PBCs) (e.g., for a one-dimension critical spin chain of total size $L$):
\begin{equation}
    S^{\text{PBC}}(L,l) = \frac{c}{3}\ln\left(\frac{L}{\pi}\sin\left[\frac{\pi l}{L}\right]\right) + c_{1}^{\text{PBC}} \, ,
\end{equation}
where $c_{1}^{\text{PBC}}$ is a non-universal constant, and $c$ is the central charge of the underlying CFT. 
The central charge serves as a fingerprint for classifying the bulk universality class of quantum phase transitions and has garnered extensive theoretical and experimental attention~\cite{Calabrese_2004,islam2015measuring}.

However, in the presence of a boundary, an intriguing degree of freedom emerges where the bulk of the system can remain at a critical point, while the boundary may flow between different fixed points under boundary renormalization group transformations. 
The universal behavior of a critical system with a boundary, referred to as boundary (or surface) criticality, exhibits richer physics compared to bulk universality and can be effectively characterized using boundary CFT~\cite{CARDY1984514}. 
According to the observation in Ref.~\cite {affleck1991prl}, for a system with its bulk at the critical point, one can define an Affleck-Ludwig boundary entropy. 
This boundary entropy decreases under boundary renormalization group flow and equals to a number which is related to the universality class of the corresponding conformal boundary condition at the boundary fixed point. 
These boundary critical phenomena were first discussed in detail in the 1970s and have recently gained renewed attention due to their relevance in many areas of modern physics~\cite{zhang2017prl,brillax2024prb}. 
In particular, in the context of gSPT phases or topologically nontrivial quantum critical points, recent advances~\cite{yu2022prl} have unambiguously demonstrated that conformal boundary conditions (or equivalently, the boundary $g$-function) can serve as a topological invariant for topologically nontrivial conformal critical points, at least in 1+1D, encoding nontrivial topological properties that go beyond topological edge modes. 
The boundary entropy can be explicitly defined in terms of entanglement entropy under certain boundary conditions, using a one-dimensional critical spin chain with boundary $\text{b}$ as an example: 
\begin{equation} 
    S^{\text{b}}(L,l) = \frac{c}{6}\ln\left(\frac{2L}{\pi}\sin\left[\frac{\pi l}{L}\right]\right) + \frac{c^{\text{PBC}}_{1}}{2} + \ln g + \mathbb{G}_{b}(l/L) \, , 
\end{equation} 
where $g$ is the boundary $g$-function, which fully determines the boundary universality class and plays a role analogous to the central charge for bulk universality. 
The term $\mathbb{G}_{b}$ represents a scaling function, which is generally challenging to determine analytically. 
Consequently, extracting the boundary $ g $-function from the entanglement entropy is typically difficult, particularly in experimental settings where scalable measurements are even more challenging to achieve. 
To avoid potential confusion, we emphasize the distinction between two types of boundary conditions discussed throughout this paper: \textbf{physical boundary conditions} and \textbf{conformal boundary conditions}. 
The physical boundary condition refers to specific operations imposed on the boundary of the ultraviolet (or microscopic) lattice system, and we refer to it in the main text simply as "boundary condition". In contrast, the conformal boundary condition corresponds to the boundary fixed point that emerges under the boundary renormalization group flow in the infrared limit. 
Importantly, these two types of boundary conditions are generally not directly related.

\subsection{Extract the boundary $g$-function from the wavefunction overlaps} \label{sec:g_overlap}

To address the scaling challenges associated with entanglement entropy to extract the boundary $g$-function, recent theoretical progress~\cite{Zou2022PRB,Zhou2024SciPost} have introduced a wavefunction overlap method. This approach offers a direct means to determine the value of the boundary $g$-function, eliminating the need to extract it indirectly through fitting the entanglement entropy. Thanks to the operator-state correspondence in CFT, expressions in terms of primary operators can be translated into the form involving the corresponding states. 
More specifically, as proofed in Ref.~\cite{Zhou2024SciPost}, the overlap between two low-lying eigenstates of the Hamiltonian with different defect deformations can be related to universal defect data, such as the defect $g$-function, the scaling dimension of the defect change operator $\Delta_{\alpha}^{ab}$, and the four-point correlation function for defect change operators. Finally, by choosing suitable eigenstates, the universal data can be extracted from the ratio of different wavefunction overlaps by canceling out the unwanted parts. In particular, the $g$-function can be obtained by taking the ratio 
\begin{equation}
    \label{eq:g_function}
    g_{a}^{\rm Def} = \Bigg( \frac{\langle{\phi_{0}^{00}}|{\phi_{0}^{a0}}\rangle}{\langle{\phi_{0}^{aa}}|{\phi_{0}^{a0}}\rangle} \Bigg)^{2} \,,
\end{equation}
where $|{\phi_{\alpha}^{ab}}\rangle$ is the eigenstate of the Hamiltonian with defect deformations
\begin{equation}
    H^{ab} = H + \lambda \mathcal{O}_{L-1,0}^{a} + \lambda' \mathcal{O}_{L/2-1,L/2}^{b} \,,
\end{equation}
which corresponds to the primary operator $\phi_{\alpha}^{ab}$. 
The superscript $a$ denotes the defect type ($a = 0$ means the trivial defect) while the subscript $\alpha$ labels the primary operator ($\alpha = 0$ means the lowest primary operator corresponding to the ground state of the defect Hamiltonian) and the perturbation $\mathcal{O}^{a}_{j,j+1}$ is the lattice realization of the defect type $a$ at sites $j$ and $j+1$. 

It is noted that the $g$-function, $g_{a}^{\rm Def}$, in Eq.~\eqref{eq:g_function} is the defect $g$-function. 
Actually, the boundary is a special type of line defect and the boundary $g$-function is the square root of its defect correspondence~\cite{OSHIKAWA1997533}. 
In the following, we use $g_{a}$ to denote the boundary $g$-function for the boundary condition $a$ and the square in Eq.~\eqref{eq:g_function} should be removed in its calculations.

At last, we note that the expression Eq.~\eqref{eq:g_function} can be generalized to low-lying excited states corresponding to defect primary operators and the ratio of the wavefunction overlaps is
\begin{equation}
    \frac{\langle \phi^{00}_{\beta} | \phi^{a0}_{\alpha} \rangle}{\langle \phi^{aa}_{\gamma} | \phi^{a0}_{\delta} \rangle} = g_{a} \frac{C^{0a0}_{0\alpha \beta}}{C^{a0a}_{\delta 0 \gamma}} \,,
\end{equation}
where $C^{0a0}_{0\alpha\beta}$ and $C^{a0a}_{\delta 0 \gamma}$ are defect operator product expansion (OPE) coefficients. 
It means that the ratio of overlaps between different excited states is proportional to the boundary $g$-function, with the proportionality constant determined by the ratio of the corresponding OPE coefficients. 
In principle, if the OPE coefficients of the underlying CFT are known, the boundary $g$-function can be extracted using the wavefunction overlap method, even for excited states. 
In our experiments, which corresponds to the Ising CFT, the $\mathbb{Z}_{2}$-odd sector $\sigma$ under the periodic boundary condition is projected out as part of an error mitigation strategy because the correct ground state is in the $\mathbb{Z}_{2}$-even sector. 
Under this condition, the ratio between the relevant OPE coefficients all equal to $1$~\cite{Zou2022PRB}, leading to a direct equivalence between the wavefunction overlap of low-lying excited states and the boundary $g$-function. 
This simple argument provides an intuitive explanation for why an accurate boundary $g$-function can be obtained from low-lying critical states. 
A more detailed and systematic investigation of this significant and intriguing question for general cases is left for future work.

\subsection{Universal entanglement spectrum at topologically nontrivial critical states}

The concept of the entanglement spectrum is a powerful tool in the study of topological phases of matter and SPT physics. 
As first noticed by Li and Haldane in their seminal work~\cite{Li2008PRL}, the entanglement spectrum of a reduced density matrix is believed to encode additional information on the existence of the degenerate edge modes at the boundary of a gapped topological phase, which cannot be captured by the entanglement entropy. 
However, a remaining question is whether this observation still holds in gapless topological phases. 
This question has been addressed in a recent work~\cite{yu2024prl,zhang2024pra}, at least in the context of one-dimensional gapless SPTs. 
Through extensive numerical calculations and conformal mappings, it has been found that the entanglement spectrum not only captures the topological edge degeneracy but also the operator content of the boundary CFT. 
Therefore, to experimentally observe the nontrivial edge degeneracy of the critical cluster-Ising chain, one can probe the entanglement spectrum of a contiguous interval of the periodic critical chain; 
the two-fold degeneracy in the low-lying entanglement spectrum reflects the nontrivial edge states. 
In this work, the bulk entanglement spectrum is obtained from the prepared state via the Entanglement Hamiltonian Tomograph (EHT) method, which can be performed efficiently on digital quantum simulation platforms.

\begin{figure*}[tb] 
\begin{center}
    \includegraphics[width=0.65\textwidth]{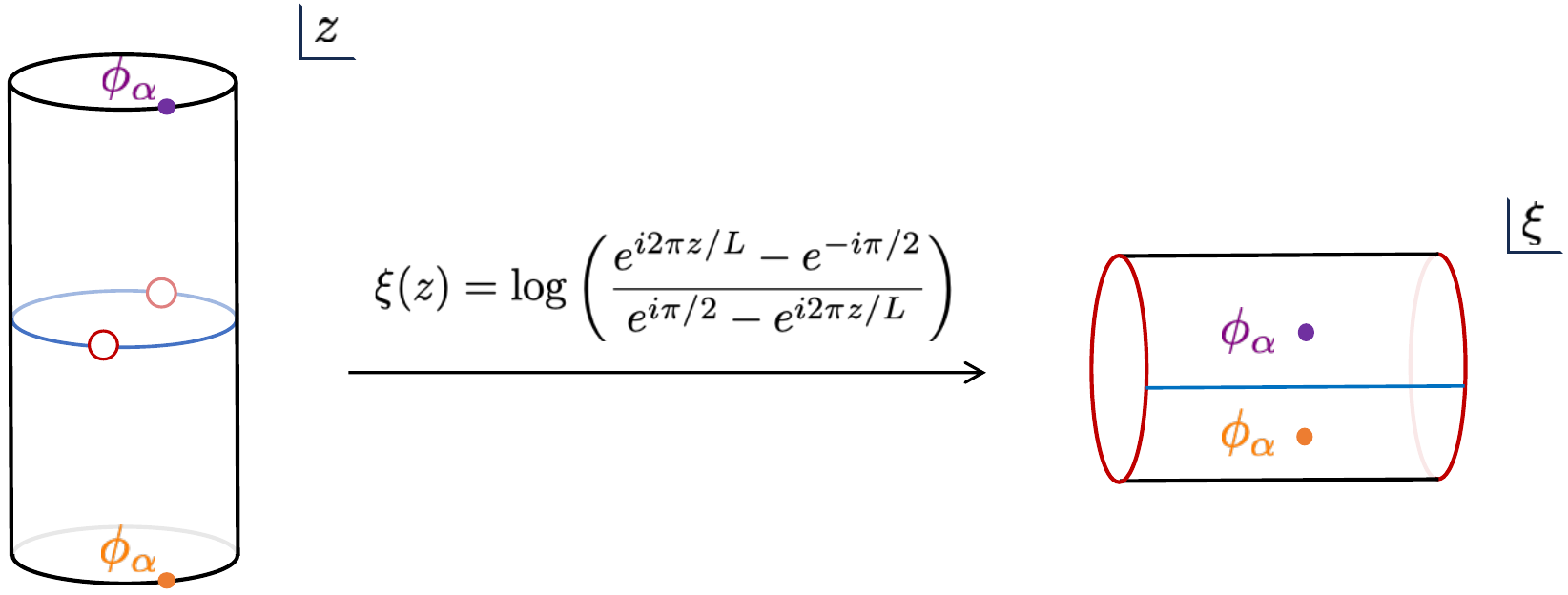}
    \caption{Explanation for the double degenerate of the entanglement spectrum calculated from the low-lying quantum critical state. The red circle represents the UV cutoff induced by the entanglement cut and $\phi_{\alpha}$ is the primary operator inserted at the infinity past/future imaginary time, which corresponds to the eigenstate $|\phi_{\alpha}\rangle$ of the periodic Hamiltonian. Via the conformal mapping $\xi(z)$, the primary operators are mapped into local operators in the bulk, which does not influence the boundary in the imaginary-time evolution.} 
    \label{fig_si_conformal}
\end{center} 
\end{figure*}

To provide an intuitive explanation for why the low-lying quantum critical state generated in experiments can effectively exhibit topological degeneracy in the bulk entanglement spectrum, we begin with the state-operator correspondence in CFT. According to this principle, a low-lying excited state can be interpreted as an operator inserted at infinity, as illustrated in Fig.~\ref{fig_si_conformal}. By applying a series of conformal transformations, the reduced density matrix of the low-lying quantum critical state can be mapped onto a cylinder. In this context, the excited states correspond to bulk local operators $\phi_{\alpha}$ in imaginary time. However, the critical cluster Ising model is known to host two decoupled fractionalized edge states in the Majorana representation~\cite{verresen2018prl}. Consequently, the imaginary-time evolution of the bulk local operator does not influence the edge states, thereby preserving the double degeneracy in the entanglement spectrum. The more detailed proof for general cases can be left as future work.

\section{Details of the density matrix renormalization group method}

To better understand the prepared state, we need to analyze its overlap with the low-lying excited states of the target many-body Hamiltonian.
For this purpose, we employ the Density Matrix Renormalization Group (DMRG) method based on the Matrix Product State (MPS) formalism to solve the first several low-lying eigenstates of the cluster-Ising model under different boundary conditions.

By exploiting the MPS representation, a quantum state can be expressed in a compact form:
\begin{equation}
    \ket{\psi} = \sum_{\{\sigma_{i}\}} M^{\sigma_{1}}_{1,a_{1}} \cdots M^{\sigma_{i}}_{a_{i-1},a_{i}} \cdots M^{\sigma_{L}}_{a_{L-1},1} \ket{\sigma_{1}\cdots\sigma_{L}} \,,
\end{equation}
where $\sigma_{i} =\, \uparrow, \downarrow$ is a local basis of the qubit on site $i$, and $M^{\sigma_{i}}$ are matrices with appropriate dimensions so that their multiplication leads to a scalar, namely, the wavefunction $\langle \sigma_{1} \cdots \sigma_{L} | \psi \rangle$. 
The ground state of the Hamiltonian can be obtained by optimizing the local matrices according to the variational principle. 
Once the expectation value of the Hamiltonian with respect to the MPS has reached the convergence criterion, the optimization process would stop and the final MPS can be used to represent the true ground state faithfully. 
The dimension of the bond index $\dim(a_{i})$ is a tunable parameter called bond dimension which controls the representation ability of the MPS. 
In practical simulations, we have chosen a sufficiently large bond dimension up to $1024$ to ensure the accuracy of the optimized MPS.

To access the excited eigenstates, we adopt a standard strategy by introducing an energy penalty for all the already obtained eigenstates. 
Assume that we have calculated the first $m$ low-lying eigenstates $\{ \ket{\psi_{k}} \}_{k=1}^{m}$ of the Hamiltonian, to obtain the $(m+1)$th eigenstate, we consider a modified Hamiltonian:
\begin{equation}
	H_{m+1} = H + \lambda \sum_{k=1}^{m} \ket{\psi_{k}} \! \bra{\psi_{k}} \,,
\end{equation}
where $\lambda$ is the penalty strength which should be large enough to shift the energies of the first $m$ eigenstates higher than the energy of the $(m+1)$th eigenstate ($\lambda = 50$ in our work).
Based on the modified Hamiltonian $H_{m+1}$, the $(m+1)$th eigenstate of $H$ can be accessed by running a conventional DMRG calculation. 
In this way, the first several low-lying excited states of the Hamiltonian can be systematically computed one by one. 

In the present work, the DMRG and MPS simulations are performed via ITensor~\cite{ITensor2022} and Quimb~\cite{Quimb2018} packages.

\section{Methods for Detection of Critical Topological Properties}
In this section, we detail the experimental methods used to detect the topological properties of the critical cluster Ising model. 
We devise an efficient state preparation algorithm for the cluster Ising model that prioritizes maximizing the overlap with the ground state. Additionally, we introduce a symmetry projection method to extract the boundary $g$-function. Finally, we utilize entanglement Hamiltonian tomography to probe the two-fold degeneracy of the entanglement spectrum. In this paper, quantum circuit simulations of small system sizes ($L \leq 20$) are performed via a state vector simulator from Pennylane~\cite{bergholm2022pennylaneautomaticdifferentiationhybrid} with Pytorch~\cite{NEURIPS2019_bdbca288} backend and simulations of large system sizes ($L>20$) are performed via an MPS simulator from Tensorcircuit~\cite{Zhang2023tensorcircuit} with Tensorflow~\cite{tensorflow2015-whitepaper} backend.

\subsection{Low-lying State Preparation} \label{sec:state_prep}
In this paper, we devise an efficient way to prepare the low-lying state of the CI model based on classically optimized variational quantum circuits and parameter extrapolation. 
We select a hardware-efficient ansatz containing five blocks of variational circuits, with each block containing a layer of parameterized $Y(\theta)$ rotation gates followed by a layer of CZ gates, as shown in Fig.~2\textbf{a} of the main text.
This construction constrains the trial states to have real-valued amplitudes.
The CZ gate layers in odd (even) blocks have identical structures and follow the spin connection topology of the Hamiltonian, respecting the translational symmetry of the system in the bulk.
For PBC, the even CZ gate layers contain gates acting on the first and last qubits to keep the translational symmetry on the whole chain. 

We further introduce an inductive bias into the circuit design by modeling the variational parameters as a site-dependent function. 
For PBC, the parameters in the same layer are set to be identical, i.e. $\theta_{i,j}^\text{PBC}=c_j^\text{PBC}$ with $i$ ($j$) indexing the qubit (block) number.
For OBC, due to the presence of boundaries, we introduce a perturbation that follows the power-law decay of the rotation angles from the boundary in each layer:
$\theta_{i,j}^\text{OBC}=a_j^\text{OBC}(D_i+1)^{b_j^\text{OBC}}+c_j^\text{OBC}$,
where $D_i$ represents the distance of the $i^{\text{th}}$ qubit from the boundary. 
We first optimize all parameters by minimizing the energy of the 8-qubit system.
The power parameter $b_j^\text{OBC}$ is directly determined, and other parameters ($a_j^\text{OBC}$, $c_j^\text{OBC}$, and $c_j^\text{PBC}$) are further optimized to maximize the overlaps between the low-lying states and target states.
Specifically, for OBC, we apply the modification $a_j^\text{OBC}\rightarrow a_j^\text{OBC}+\epsilon_j$ and $c_j^\text{OBC}\rightarrow c_j^\text{OBC}-\epsilon_j$, and optimize $\epsilon_j$ by minimizing the loss function
$\mathcal{L}^\text{OBC}=1-|\langle\Psi_v^\text{OBC}|\psi^\text{OBC}_\text{gs.}\rangle|$, where $\psi^\text{OBC}_{\text{gs.}}$ denotes the ground state. 
For PBC, due to the limited expressivity of the ansatz, the loss function is selected to be
$\mathcal{L}^\text{PBC}=1-\sqrt{{|\langle\Psi_v^\text{PBC}|\psi^\text{PBC}_\text{gs.}\rangle|}^2+{|\langle\Psi_v^\text{PBC}|\psi^\text{PBC}_\text{1st}\rangle|}^2}$,
where $\psi^\text{PBC}_{\text{1st}}$ is the first excited state. 
After obtaining $\epsilon_j$ and $c_j^\text{PBC}$ for systems with sizes $L = 8, 10, 12, 14$, and $16$, we apply extrapolation to obtain the two parameters for larger system sizes.
The extrapolating functions take the form of $f(L)=a(L+d)^{b}+c$ for both parameters, which agree well with the directly optimized ones for $L=18$ and $20$, respectively, as displayed in Fig.~\ref{fig_si_pt}.
The resulting parameters of the third circuit block for OBC and PBC are displayed in Fig.~2\textbf{b} in the main text, and the parameters of the lefting circuit blocks are shown in Fig.~\ref{fig_si_cp}. 

\begin{figure*}[!t] 
\begin{center}
    \includegraphics[width=\textwidth]{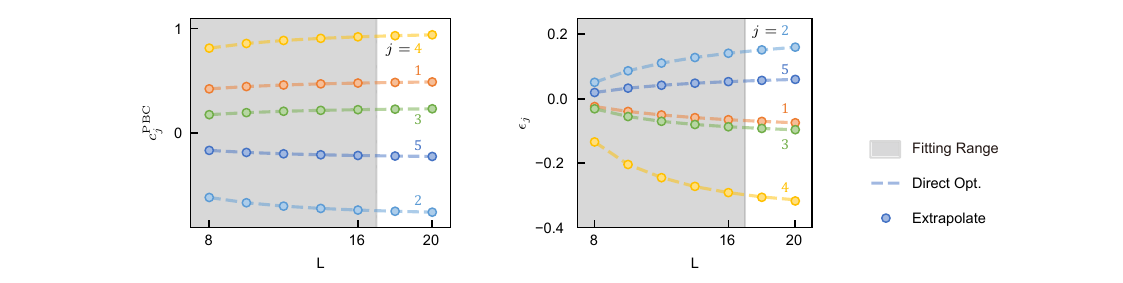}
        
    \caption{\textbf{Verification of the extrapolating functions.}
    The dotted lines indicate parameters optimized directly without using energy-initialized parameters. Markers indicate the fitted and extrapolated parameters.
    } \label{fig_si_pt}
\end{center} 
\end{figure*}

\begin{figure*}[!t] 
\begin{center}
    \includegraphics[width=\textwidth]{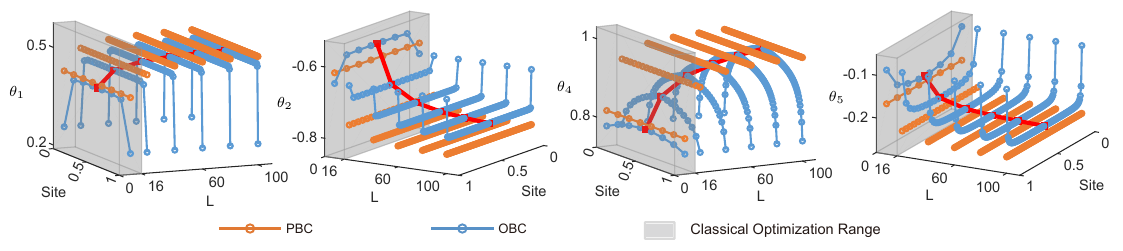}
        
    \caption{\textbf{Single-qubit rotation angles for other layers of the circuit.}} \label{fig_si_cp}
\end{center} 
\end{figure*}

\subsection{Boundary $g$-Function Measurement}

The original proposal for calculating the boundary $g$-function is based on the overlaps between the ground states of the system under different boundary conditions.
As explained in Sec.~\ref{sec:g_overlap}, the method can be generalized to excited eigenstates for Ising CFT by constraining the states to the $\mathbb{Z}_{2}$-even sector.
In our experiment, we take a step further to use low-lying states to calculate the boundary $g$-function.

A key step involves projecting the low-lying state under PBC to the $\mathbb{Z}_{2}$-even sector with the projector $\mathcal{P}=(I+O_X)/2$, which naturally excludes the first excited state and leaves the projected state dominated by the ground state.
Specifically, we can express the PBC low-lying state as the superposition of two wavefunctions in the $\mathbb{Z}_{2}$-even and -odd sectors: $|\Psi_v^{00}\rangle=\alpha_+|\Psi_v^{00,+}\rangle + \alpha_-|\Psi_v^{00,-}\rangle$, with $O_X|\Psi_v^{00,\pm}\rangle = \pm|\Psi_v^{00,\pm}\rangle$.
After some algebra, we have $|\Psi_v^{00,+}\rangle=\frac{\mathcal{P}}{\alpha_+}|\Psi_v^{00}\rangle$, with $\alpha_+^2=\langle\Psi_v^{00}|\mathcal{P}|\Psi_{v}^{00}\rangle$.
Thus the $g$-function can be calculated as
\begin{align}
g=\sum_{a\in\{\uparrow, \downarrow\}}\left|\frac{\langle\Psi_v^{a0}|\Psi_{v}^{00,+}\rangle}{\langle\Psi_v^{a0}|\Psi_v^{aa}\rangle}\right|=\sum_{a\in\{\uparrow, \downarrow\}}\frac{1}{\sqrt{\langle\Psi_v^{00}|\mathcal{P}|\Psi_{v}^{00}\rangle}}\left|\frac{\langle\Psi_v^{a0}|\mathcal{P}|\Psi_{v}^{00}\rangle}{\langle\Psi_v^{a0}|\Psi_v^{aa}\rangle}\right| \,.\label{eq:g_explicit}
\end{align}

The experimental measurement of Eq.~\ref{eq:g_explicit} is possible due to the fact that the variationally prepared low-lying states are all real-valued. As a result, the problem can be reduced to the general problem of measuring $|\langle\Psi_0|O|\Psi_1\rangle|$, with $O=I$ or $O_X$.
In our experiment, the state $|\Psi_{0 (1)}\rangle$ is prepared by applying the corresponding variational circuit $U_{0 (1)}$ on the initial state $|0^{\otimes L}\rangle$, i.e., $|\Psi_{0 (1)}\rangle=U_{0 (1)}|0^{\otimes L}\rangle$.
Thus we have $|\langle\Psi_0|O|\Psi_1\rangle|=|\langle 0^{\otimes L}|U_0^\dagger O U_1|0^{\otimes L}\rangle|$, which can be measured in experiment by applying the circuits $U_1$, $O$, and the reverse circuit $U_0^\dagger$ sequentially on the initial state $|0^{\otimes L}\rangle$, and detecting the probability of the final state being in $|0^{\otimes L}\rangle$.

\subsection{Entanglement Hamiltonian Tomography for Entanglement Spectrum Detection}
In this paper, we use entanglement Hamiltonian tomography (EHT) to extract the ground state entanglement spectrum from the experimental data. We parameterize the reduced density matrix of a subsystem $A$ as $\rho_A(\beta)=e^{-\widetilde{H}_A}$, with the entanglement Hamiltonian $\widetilde{H}_A$ defined in Eq.~2 in the main text, and $\beta=[\beta^{ZZ}; \beta^{ZXZ}; \beta^{X}]$.

After preparing the low-lying state, we apply random basis rotations $U^{(\alpha)}=\otimes_{j\in A}u_j^{(\alpha)}$, where $u_j^{(\alpha)}$ are single-qubit gates that randomly rotate the $j^\text{th}$ qubit from $x$-,$y$-,$z$- basis to the $z$-basis, and measure the bitstring probabilities in the $z$-basis. Specifically, we perform 200 different measurement settings $\alpha$ and collect 3000 measurements for each setting. Denoting the probability of measuring the bitstring $\mathbf{s}$ under the $\alpha^\text{th}$ measurement setting by $P_{\mathbf{s}}^{(\alpha)}$, we minimize the following loss function on the classical computer:

\begin{align}
\mathcal{L}^\text{EHT} & = \sum_{\alpha} \sum_{\mathbf{s}}\left[P_{\mathbf{s}}^{(\alpha)}-\operatorname{Tr}\left(\rho_{A}(\beta) U^{(\alpha)}|\mathbf{s}\rangle\langle\mathbf{s}| U^{(\alpha) \dagger}\right)\right]^{2} \,.
\label{eq:EHT_loss}
\end{align}

In practice, we use the Lanczos algorithm to get the smallest 12 eigenvalues $\lambda_{i}$ and the corresponding eigenvectors $|\Phi_A^i\rangle$ of $\widetilde{H}_A$.
This allows us to simplify the second term on the right-hand side of Eq.~\ref{eq:EHT_loss} to $\sum_{i=1}^{12} \mathrm{e}^{-\lambda_i}\left|\left\langle\Phi_{A}^{i}\right| U^{(\alpha)}\left| \mathbf{s}\right\rangle\right|^{2}$, enabling a more efficient classical calculation.
This optimization is performed for subsystem sizes ranging from 5 to 12, with 100 different subsystems analyzed for each subsystem size in the 1D chain. For each subsystem, we randomly initialize and optimize 3 groups of $\beta$s, and select the $\beta$ with the lowest loss. 

\section{Experimental Details}

\subsection{Device Information}

The experiment is performed on a superconducting quantum chip containing 125 qubits and 218 couplers, from which spin chains with various sizes can be constructed, as illustrated in Fig.~\ref{fig:topo}.

\begin{figure}[h] 
    \begin{center}
    \includegraphics[width=0.8\columnwidth]{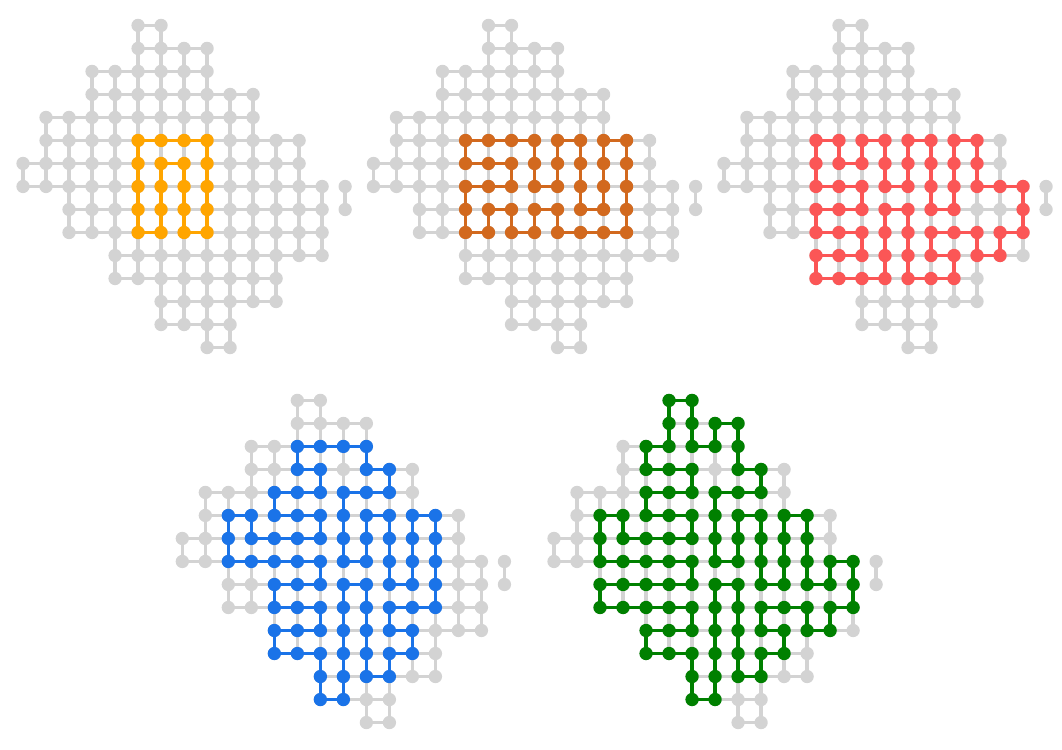}
    
    \caption{\textbf{Qubit layout details.} 
    We construct 1D qubit chains with 20, 40, 60, 80, and 100 qubits out of the 125-qubit superconducting processor.}
    \label{fig:topo}
    \end{center}
\end{figure}

The qubits have a mean energy relaxation time $T_1$ of 64.6 $\mu$s and dephasing time (obtained by spin Echo) $T_2$ of 19.1 $\mu$s, which are measured at the qubit idling frequencies.
The single- and two-qubit gates are implemented following the procedure outlined in \cite{Xu2024Fib}, with gate lengths of 20 and 32 ns, respectively.
The qubit measurements are optimized using the method described in \cite{wang2024nonlocality}.
The distributions of gate and measurement error rates are presented in Fig.~\ref{fig:gate_fidelity}, with dashed lines indicating the median values. 
Single- and two-qubit gate errors are estimated with simultaneous cross-entropy benchmarking technique.

\begin{figure}[h] 
    \begin{center}
    \includegraphics[width=0.5\columnwidth]{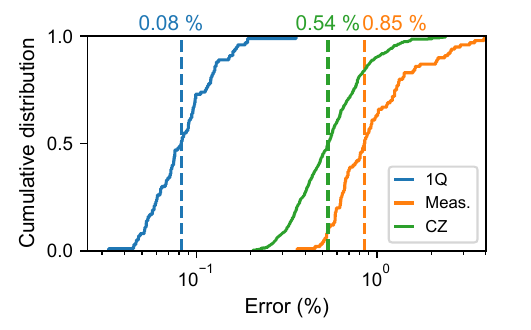}
    
    \caption{\textbf{Cumulative distributions of gate Pauli errors and measurement error.} The dashed lines indicate the median values.
    We consider every CZ gate used in our experiments into the distribution, with some couplers used multiple times. 
    The measurement error is calculated from the average error of $\ket{0}$ and $\ket{1}$.
    }
    \label{fig:gate_fidelity}
    \end{center}
\end{figure}

\subsection{Error Mitigation}

In this part, we outline the error mitigation strategies employed in our experiments. 
Specifically, we combine Zero Noise Extrapolation (ZNE) with Pauli twirling (PT) and circuit simplification for energy measurements and use PT and circuit simplification for boundary $g$-function measurements.

\subsubsection{Zero Noise Extrapolation}

To mitigate the effects of noise in the quantum circuits, we apply Zero Noise Extrapolation (ZNE) in conjunction with random Pauli twirls. We estimate the error-free outcome by running unitary-equivalent quantum circuits at different noise levels. Due to the regular structure of our state preparation circuit, it is particularly well-suited for applying ZNE with unitary folding~\cite{LiYing2017}, which increases the noise level of the circuit.
We implement Pauli twirling (PT) at each noise level to transfer the coherent errors into Pauli errors, which involves randomly applying a sequence of Pauli operations to the quantum system and averaging the results over multiple randomized circuits. We observe that the averaged outcomes become stable when the number of random Pauli-twirled circuits reaches 25.
We then extrapolate the circuit's outcome as the noise approaches zero with a linear model. This enables us to estimate the ideal, noise-free results.

In Fig.~\ref{fig:ZNE}, we present the results of the energy measurement experiment and the error-mitigated outcomes for the 100-qubit PBC ground state. The experimental state preparation circuit consists of 4 CZ gate layers and 5 single-qubit gate layers. 
To increase the noise levels, we randomly select and fold different numbers of CZ gate layers as well as the adjacent single-qubit gate layers.
For example, the noise scale factor would be three if we fold each CZ layer exactly once.
The error bars are computed using bootstrapping, where we resample shots uniformly with replacements from the pool of shots for all 25 Pauli-twirled circuits. This procedure generates 100 mitigated energy expectation values, from which we calculate the mean and standard deviation.

We tried to use both linear and exponential extrapolation to extract the error-mitigated value.
While the exponential extrapolation could sometimes yield outcomes with less errors, we found a larger deviation compared with the results obtained based on linear extrapolation.
Besides, we also observed that the exponential model is less stable than the linear model.
Therefore, we adopted linear extrapolation in our experiment.

\begin{figure}[h] 
    \begin{center}
    \includegraphics[width=0.5\columnwidth]{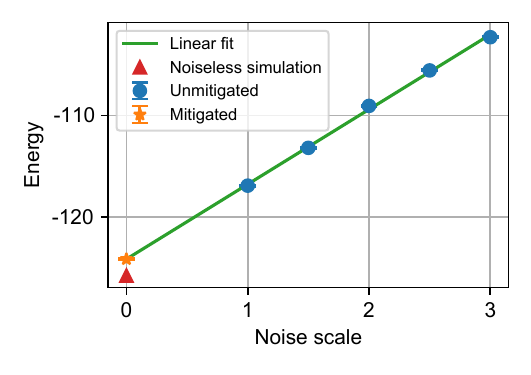}
    
    \caption{\textbf{Energy measurement with zero noise extrapolation.} 
    We display the measured energies of a 100-qubit PBC ground state with different noise scale factors $( F \in \{1, 1.5, 2, 2.5, 3\} )$. 
    For each noise scale, we add Pauli twirls in 25 randomly folded circuits.
    Error bars indicate $68\%$ confidence interval, obtained from bootstrapping 100 configurations.
    }
    \label{fig:ZNE}
\end{center} \end{figure}

\subsubsection{Circuit Simplification}
During the measurement of state energies and overlaps, the circuit can be further simplified according to the commutation relations between CZ gates and single-qubit Pauli operators, as depicted in Fig.~\ref{fig:circuit_simplify}.

\begin{figure}[h] 
    \begin{center}
    \includegraphics[width=0.8\columnwidth]{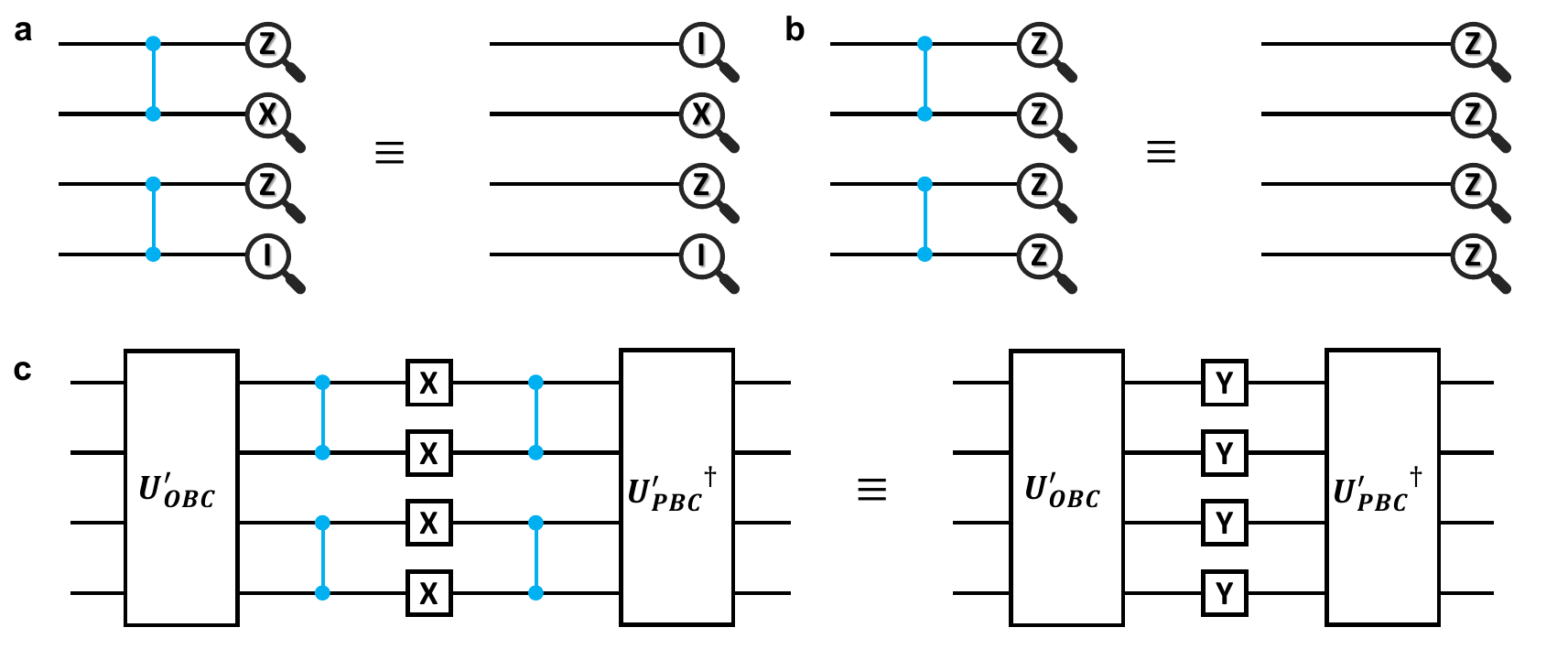}
    
    \caption{\textbf{Circuit simplification for energy and overlap measurement.}
    \textbf{a} and \textbf{b}, In the energy measurement circuit, the final CZ gate layer is eliminated by modifying the observables according to Eq.~\ref{eq:circuit_simplify}.
    \textbf{c}, In the overlap measurement circuit, the Pauli $X$ gates, which are sandwiched between two CZ gate layers in the middle of the circuit, can be simplified to a single layer of Pauli $Y$ gates.}
    \label{fig:circuit_simplify}
\end{center} \end{figure}

Specifically, rewriting the unitary of the state preparation circuit as $U=U_\text{CZ}U'$, where $U_\text{CZ}$ represents the unitary of the last CZ layer, and $U'$ denotes the circuit unitary excluding the last CZ layer, we have $|\Psi\rangle=U_{\text{CZ}}U'|0\rangle^{\otimes L}$.
The expectation value of an observable $O$ can be written as
\begin{align}
\langle\Psi|O|\Psi\rangle=\langle0|^{\otimes L}U'^\dagger U_{\text{CZ}} O U_{\text{CZ}} U' |0\rangle^{\otimes L}=\langle0|^{\otimes L}U'^\dagger O^{'} U' |0\rangle^{\otimes L} \,,
\label{eq:circuit_simplify}
\end{align}
where $O'=U_\text{CZ}O U_{\text{CZ}}$. Since the observable $O$ in our experiment is a Pauli operator, $O'$ remains a Pauli operator, enabling the removing of CZ layer by substituting $O$.

\subsubsection{Intrinsic Noise Resistance}
The boundary $g$-function measurement and EHT-based entanglement spectrum detection are naturally resistant to noise. For the boundary $g$-function, both the numerator and denominator overlaps will decrease under decoherent noise, but their ratio remains nearly constant. In Fig.~\ref{fig:g_noisy_sim}, We simulate the effect of CZ depolarization error on the measurement of boundary $g$-function for a system size $L=8$, and observe that the resulting $g$-function remains robust over a wide range of noise rates.

\begin{figure}[h] 
    \begin{center}
    \includegraphics[width=0.5\columnwidth]{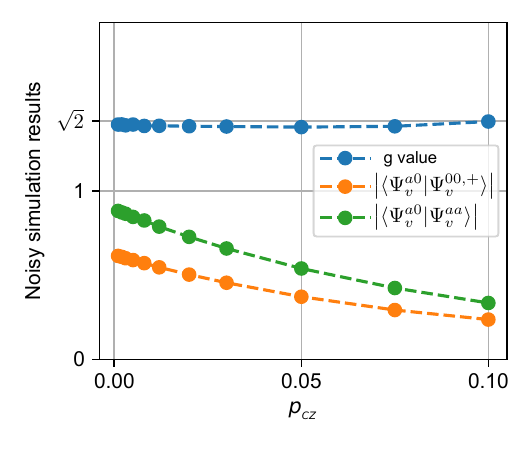}
    
    \caption{\textbf{Robustness of $g$-function measurement.} We perform noisy circuit simulations on 8 qubits with experimentally measured $T_1$ and $T_2$. We add an additional CZ depolarization channel with a controlled depolarizing rate, denoted by $p_{_{CZ}}$. While the overlap decreases as the depolarizing rate increases, the $g$-function value remains almost unaffected, demonstrating its robustness.}
    \label{fig:g_noisy_sim}
\end{center} \end{figure}

To demonstrate the robustness of EHT to noises, we first extract the noiseless density matrix $\rho$ of an eight-qubit subsystem from the 100-qubit ground state with PBC, which is obtained using DMRG.
We then introduce a depolarizing channel to the system, obtaining the noisy density matrix $\rho_{n} = (1-p)\rho + p I/d$, where $p$ is the depolarizing error rate and $d=2^8$.
As shown in Fig.~\ref{fig:EHT_noisy_sim}, the fidelity between $\rho$ and $\rho_n$ decreases with the increase of the depolarizing error rate as expected.
In contrast, when we apply EHT on the eight qubits with the loss function defined in Eq.~\ref{eq:EHT_loss} and random 200 Pauli bases, we find that the fidelities of the resulting density matrices $\rho_\text{EHT}$ to $\rho$ are much less affected by the depolarizing noises.

\begin{figure}[h] 
    \begin{center}
    \includegraphics[width=0.5\columnwidth]{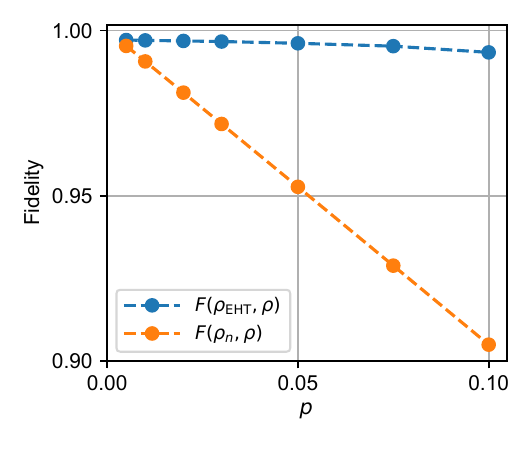}
    
    \caption{\textbf{EHT noisy simulation.} We compare the fidelity between $\rho$ 
    and $\rho_\text{EHT}$ 
    , as well as $\rho_n$ 
    with different depolarizing rates $p$.
    The results demonstrate that the EHT learning protocol is robust to depolarizing noise.}
    \label{fig:EHT_noisy_sim}
\end{center} \end{figure}

\end{widetext}

\end{document}